\documentclass[twocolumn,preprintnumbers,amsmath,amssymb]{revtex4}

\usepackage{epsfig}
\usepackage{bm}

\begin{document}


\title{Global monopole surrounded by quintessence-like matter }

\author{Xin-zhou Li}\email{kychz@shnu.edu.cn}
\author{Ping Xi}
\author{Xiang-hua Zhai}
\affiliation{Shanghai United Center for Astrophysics(SUCA),\\
 Shanghai Normal University, 100 Guilin Road, Shanghai 200234,China
}

\date{\today}

\begin{abstract}
We present new static spherically-symmetric solutions of Einstein
equations with the quintessence-like matter surrounding a global
monopole. These new solutions of the coupling scalar-Einstein
equations are more complicated, which depend on the parameter of
equation of state $-1 < w_{q} <-\frac{1}{3}$. A gravitating global
monopole produces a gravitational field of de Sitter kind outside
the core in addition to a solid angular deficit. In the $w_{q} =
-\frac{1}{3}$ case, we have proved that the solution cannot exist
since the density of quintessence-like tends to zero if $w_{q}
\rightarrow -\frac{1}{3}$. As a new feature, these monopoles have
the outer horizon depending on both Goldstone field and
quintessence-like. Since current observations constrain  $-1.14 <
w_{q} < -0.93$, new global monopoles have interesting astrophysical
applications.
\end{abstract}

\maketitle

\noindent { \bf  1. Introduction}\\
\\
\noindent \indent The phase transition in the early universe could
have produced different kinds of topological defects which have some
important implications in cosmology \cite{Shellard}. Point-like
defects which undergo spontaneous symmetry breaking arise in some
theories and they appear as monopoles. The global monopole, which
has divergent mass in flat space-time, is one of the most
interesting defects. The idea that monopoles ought to exist has
proven to be remarkably durable. Barriola and Vilenkin
\cite{Barriola} first researched the characteristic of the global
monopole in curved space-time or, equivalently, its gravitational
effects. When one considers gravity, the linearly divergent mass of
the global monopole has an effect analogous to that of a deficit
solid angle plus a tiny mass at the origin. It has been shown that
this small gravitational potential is actually repulsive \cite{Shi}.
Liebling \cite{Liebling} also showed that global monopoles without
horizon only exist for the parameter of deficit solid angle
$\epsilon < 1$, while for $\epsilon > \sqrt{3}$ no static solutions
exist at all. The configurations for $1 < \epsilon < \sqrt{3}$ were
called `` supermassive " monopoles. Further, Li and co-workers
\cite{Li} have proposed a new class of cold stars called D-stars
(defect stars). One of the most important features of such stars,
compared to Q-stars, is that the theory has monopole solutions when
the matter field is absent, which makes D-stars behave very
differently from Q-stars. Topological defects are also investigated
in Friedmann-Robertson-Walker space-time \cite{Basu}. It is shown
that the properties of global monopoles in asymptotically de
Sitter/anti-de Sitter (dS/AdS) space-time \cite{Lu} and Brans-Dicke
theory \cite{Hao} are very different from those of ordinary
monopoles. The k-field theory, in which the non-canonical kinetic
terms are introduced in the Lagrangian, has been investigated in the
inflation scenario and the cosmic coincidence problem. Another
interesting application of k-field is topological defects, called
k-defects \cite{Babichev}. Monopoles \cite{Liu} and vortices
\cite{X} of tachyon field, which are examples of k-fields coming
from string/M-theory, have also been studied. \\
\indent The huge attractive force between a global monopole $M$ and
an anti-monopole $\bar{M}$ implies that the monopole over-production
problem does not exist, because pair annihilation is very efficient.
Barriola and Vilenkin have shown that the radiative lifetime of the
pair is very short as they lose energy by Goldstone boson radiation
\cite{Barriola}. No serious attempt has been made to develop an
analytical model of the cosmological evolution of a global monopole,
so one is limited to the numerical simulations of evolution
\cite{Bennett}. In the $\sigma$-model approximation, the average
number of monopoles per horizon is $N_{H} \sim 4$. The gravitational
field of global monopoles can lead to clustering of matter, and can
later evolve into galaxies and clusters. The scale-invariant
spectrum of fluctuations has been given in \cite{Bennett}. Moreover,
one can numerically obtain the CMB anisotropy $(\delta T/T)_{rms}$
patterns \cite{Rhie}. Comparing the theoretical value to the
observed rms fluctuation, one can find the constraint of parameters
in the global monopole. Therefore, the global monopole is considered
to be relevant to structure formation in the early universe.\\
 \indent On the other hand, current observations, such as CMB
(Cosmic Microwave Background) anisotropy, SNeIa (Supernovae type Ia)
and large scale structure, converge on the fact that a spatially
homogeneous and gravitationally repulsive energy component, referred
as dark energy, accounts for about $70$ \% of the energy density of
universe. Some heuristic models that roughly describe the observable
consequences of dark energy were proposed in recent years, a number
of them stemming from fundamental physics \cite{Padmanabhan} and
other being purely phenomenological \cite{Copeland}. One possibility
is that the Universe is permeated by an energy density, constant in
time and uniform in space. In this case, the ratio $w$ of pressure
to energy density (the equation of state) is also a constant at all
time. Another possibility is that the dark energy is some kind of
dynamical fluid, so that the ratio is a function $w(a)$, where $a$
is the scale factor of universe. Whatever, $w$ should be in the
range of $-1 \leq w < -\frac{1}{3}$, which is the requirement of
cosmic acceleration. Furthermore, the Einstein equations for the
static spherically-symmetric quintessence-like surrounding a black
hole\cite{Kiselev} have been investigated. Kiselev gave
the general expression for the static spherically-symmetric energy-momentum tensor of quintessence-like matter \cite{Kiselev}.\\
\indent In this paper, we investigated new static
spherically-symmetric solutions of Einstein equations with the
 quintessence-like matter surrounding a global monopole. These new solutions of
 the coupling scalar-Einstein equations are more complicated, which depend on
 the parameter of equation of state $-1 < w_{q} < -\frac{1}{3}$. We
 show that the gravitating global monopole produces a gravitational
 field of de Sitter kind outside the core in addition to a solid angular
 deficit. Furthermore, $w_{q} = -\frac{1}{3}$ solution cannot exist
 because the density of quintessence-like tends to zero as $w_{q} \rightarrow
 -\frac{1}{3}$. The new feature is the appearance of outer horizon
 for the case of quintessence-like matter surrounding a global
 monopole. Especially, we studied the $w_{q} = -\frac{2}{3}$ case in
 detail.\\
 \\
 \noindent {\bf 2. Equations of motion}\\
 \\
\noindent \indent We shall work within a particular model in unit $c
= 1$, where a global $O(3)$ symmetry is broken down to $U(1)$. The
Lagrangian density is
\begin{equation}
\mathcal{L}=
\frac{1}{2}g^{\mu\nu}\partial_{\mu}\phi^{a}\partial_{\nu}\phi^{a}-\frac{\lambda^{2}}{4}(\phi^{a}\phi^{a}-{\sigma_{0}}^{2})^{2}.
\end{equation}
where $\phi^{a}$ is triplet of scalar fields, isovector index $a =
1, 2, 3$. The hedgehog configuration describing a global monopole is
\begin{eqnarray}
\phi^{a} = \sigma_{0}f(\tilde{r})\frac{x^{a}}{\tilde{r}},\indent
with \indent x^{a}x^{a}=\tilde{r}^{2}.
\end{eqnarray}
so that we shall actually have a monopole solution if $f \rightarrow
1$ at spatial infinity and $f \rightarrow 0$ near the origin.\\
\indent The static spherically-symmetric metric can be written as
\begin{equation}
ds^{2}=B(\tilde{r})dt^{2}-A(\tilde{r})d\tilde{r}^{2}-\tilde{r}^{2}(d\theta^{2}+\sin^2(\theta)
d\varphi^{2}).
\end{equation}
with the usual relation between the spherical coordinates
$\tilde{r}, \theta, \varphi$ and the 'Cartesian' coordinate $x^{a}$.
Introducing a dimensionless $r \equiv \sigma_{0}\tilde{r}$, from (1)
and (3), we obtain the equation of motion for $f$ as
\begin{equation}
\frac{f''}{A}+[\frac{2}{Ar}+\frac{1}{2B}(\frac{B}{A})']f'-\frac{2}{r^{2}}f-\lambda^{2}(f^{2}-1)f
= 0.
\end{equation}
where the prime denotes the derivative with respect to $r$.\\
\indent When we consider the static spherically-symmetric
quintessence-like matter surrounding a global monopole, the Einstein
equations can be written as
\begin{equation}
G_{\mu\nu} = 8 \pi G(T_{\mu\nu}+\tau_{\mu\nu}).
\end{equation}
where $T_{\mu\nu}$ is the energy-momentum tensor for the Lagrangian
(1), and $\tau_{\mu\nu}$ is the energy-momentum tensor of the
quintessence-like. $\tau_{\mu\nu}$ can be characterized by a free
parameter $w_{q}$. In the static spherically-symmetric case, the
general expression for the energy-momentum tensor of
quintessence-like matter is given by
\begin{eqnarray}
{\tau_{t}}^{t} &=& {\tau_{r}}^{r} = \tilde{\rho}_{q},\\
{\tau_{\theta}}^{\theta}&=& {\tau_{\phi}}^{\phi} = -\frac{1}{2}
\tilde{\rho}_{q}(3w_{q}+1).
\end{eqnarray}
The whole situation concerning the energy-momentum tensor
 of quintessence-like matter in static spherically-symmetric case is discussed
  in great detail \cite{Kiselev}. The Einstein equations (5) now are ready to be
 written as
\begin{widetext}
\begin{eqnarray}
&&-\frac{1}{A}(\frac{1}{r^{2}}-\frac{A'}{Ar})+\frac{1}{r^{2}}=\epsilon^{2}[\frac{f^{2}}{r^{2}}+\frac{f'^{2}}{2A}+\frac{\lambda^{2}}{4}(f^{2}-1)^{2}+\rho_{q}],\\
&&-\frac{1}{A}(\frac{1}{r^{2}}+\frac{B'}{Br})+\frac{1}{r^{2}}=\epsilon^{2}[\frac{f^{2}}{r^{2}}-\frac{f'^{2}}{2A}+\frac{\lambda^{2}}{4}(f^{2}-1)^{2}+\rho_{q}],\\
&&\frac{1}{2A}(\frac{A'}{Ar}+\frac{B'^{2}}{2B^{2}}+\frac{A'B'}{2AB}-\frac{B'}{Br}-\frac{B''}{B})=\epsilon^{2}[\frac{f'^{2}}{2A}+\frac{\lambda^{2}}{4}(f^{2}-1)^{2}-\frac{3w_{q}+1}{2}\rho_{q}].
\end{eqnarray}
\end{widetext}
where the small dimensionless parameter $\epsilon$ is defined by
\begin{equation}
\epsilon\equiv\sqrt{8\pi G{\sigma_{0}}^{2}}\approx1.03 \times
10^{-16}(\frac{\sigma_{0}}{250\textup{GeV}}).
\end{equation}
and another dimensionless parameter $\rho_{q} \equiv
\tilde{\rho_{q}}/{\sigma_{0}}^{4}$.\\
\\
 \noindent {\bf 3. The solution in simplified models}\\


\noindent \indent Before solving the coupled equations for the
metric Eqs. (8)-(10) and the scalar field Eq. (4), let us analyze an
exceedingly simplified model for the monopole configuration, just to
manifest the main features of the exact solution in a naive manner.
The monopole configuration is simplified by a pure false vacuum
inside the core, and an exactly true vacuum at the exterior
\cite{Barriola}:
\begin{equation}
f =
\begin{cases}
0&\text{if $r < \delta$},\\
1&\text{if $r > \delta$}.
\end{cases}
\end{equation}
The interior solution of the Einstein equations (8)-(10) can be
written as follows
\begin{widetext}
\begin{equation}
ds^{2}=[1-\frac{\epsilon^{2}\lambda^{2}}{12}r^{2}+\frac{\epsilon^{2}\rho_{0}}{3w_{q}}r^{-3w_{q}-1}]dt^{2}-\frac{dr^{2}}{[1-\frac{\epsilon^{2}\lambda^{2}}{12}r^{2}+\frac{\epsilon^{2}\rho_{0}}{3w_{q}}r^{-3w_{q}-1}]}-r^{2}(d\theta^{2}+\sin^{2}{\theta}d\varphi^{2}).
\end{equation}
\end{widetext}
where $\rho_{0}$ is the energy density of quintessence-like matter
at $r = 1$. The Einstein equations outside the core are solved as
follows
\begin{widetext}
\begin{equation}
ds^{2}=[1-\epsilon^{2}-\frac{2GM\sigma_{0}}{r}+\frac{\epsilon^{2}\rho_{0}}{3w_{q}}r^{-3w_{q}-1}]dt^{2}-\frac{dr^{2}}{[1-\epsilon^{2}-\frac{2GM\sigma_{0}}{r}+\frac{\epsilon^{2}\rho_{0}}{3w_{q}}r^{-3w_{q}-1}]}-r^{2}(d\theta^{2}+\sin^{2}{\theta}
d\varphi^{2}).
\end{equation}
\end{widetext}
where $\epsilon$ characterizes a deficit solid angle. We easily see
that the global monopole surrounded by quintessence-like matter
generates the outer horizon of de Sitter kind at $r = r_h$ if $-1 <
w_q < -\frac{1}{3}$. From the continuity of the metric and its first
derivatives with respect to $r$, we have
\begin{eqnarray}
\delta = \frac{2}{\lambda}, \indent M = -\frac{16\pi
\sigma_{0}}{3\lambda}.
\end{eqnarray}
Therefore, it is possible to match an interior solution to an
exterior global monopole solution, but only with a negative mass
$M$. The solution (13)-(14) is certainly not an exact one for the
system of equations (4) and (8)-(10), but is an exact solution of
Eqs. (8)-(10). However, we need to point out that the simplified
version has revealed the principal aspects for the coupled Einstein
and scalar field equations. First of all, making use of Eqs.
(8)-(10) with (6) and (7), we have
\begin{equation}
\rho_{q}=\rho_{0}r^{-3(w_{q}+1)}.
\end{equation}
where the parameter $\rho_{0} > 0$ corresponds to the density of
energy  of the quintessence-like matter $\rho_{q}$ being positive.
For $-1 < w_{q} < -\frac{1}{3}$, $\rho_{q}$ is divergent at origin
which looks surprising at first glance. There is, however, no
contradiction. In fact, one should consider that the energy $E_{q}$
of quintessence-like matter up to a distance $R$ away from the
origin
\begin{equation}
E_{q} = 4\pi \sigma_{0}\int_{0}^{R}r^{2}\rho_{q}(r)dr=-\frac{4\pi
\sigma_{0}\rho_{0}}{3w_{q}}R^{-3w_{q}}.
\end{equation}
which is indeed finite. Secondly, the negative value (15) for $M$ is
not in conflict with Birkoff's theorem. As is known to all,
Birkoff's theorem affirms that the only static spherically-symmetric
vacuum solution of Einstein equations is the Schwarzschild metric,
and that the parameter $M$ in the solution is determined by the
integral of $T_t^{t}+\tau_t^{t}$ along the source. Indeed, when $R
> \delta$, we have, in the simplified model (12),
\begin{eqnarray}
&&4\pi
\sigma_{0}[\int_0^{\delta}(\frac{\lambda^{2}r^{2}}{4})+\int_\delta^{R}1+\int_0^{R}(\rho_{q}r^{2})]dr\nonumber\\
&&=4\pi \sigma_{0}R-\frac{4\pi
\sigma_{0}\rho_{0}}{3w_{q}}R^{-3w_{q}}-\frac{16\pi
\sigma_{0}}{3\lambda}.
\end{eqnarray}
This quantity is really positive, and the negative constant term is
just the same negative effective mass of Eq. (15). As Barriola and
Vilenkin have indicated that the linear term can be understood as a
kind of deficit solid angle, quite like the gravitational effects of
other topological defects, such as cosmic strings and domain walls
\cite{Barriola}.\\
\\
 \noindent {\bf 4. The solutions for coupled Einstein and scalar field equations}\\


\noindent{\it 4.1. Effective mass}\\


\noindent \indent Now our attention transfers from the simplified
version to
 the exact solution for the system to rigorously confirm the
 features which has been born of simplified one. For convenience, we
 choose the ans\"{a}tz as follows
 \begin{equation}
 A^{-1}(r) =
 1-\epsilon^{2}+\frac{\epsilon^{2}\rho_{0}}{3w_{q}}r^{-3w_{q}-1}-\frac{2G\sigma_{0}M_{A}(r)}{r}.
 \end{equation}
 where $-1 < w_{q} < -\frac{1}{3}$. Note that for $\rho_{0} = 0$,
 the existence of solutions without horizon is restricted by $\epsilon <
 1$ \cite{Rhie}. Eqs. (8)-(10) can be formally solved by
 \begin{equation}
 A^{-1}(r) =
 1-\frac{\epsilon^{2}}{r}\int_0^{r}[\frac{f^{2}}{r^2}+\frac{f'^{2}}{2A}+\frac{\lambda^2}{4}(f^2-1)^2+\rho_{q}]r^{2}dr.
 \end{equation}
 and
 \begin{equation}
 B(r) = \frac{1}{A(r)}\exp{[\epsilon^{2}\int_\infty^{r}f'^{2}rdr]}.
 \end{equation}
 As $r \rightarrow \infty$, we have $B(r) = A^{-1}(r)$. From Eqs.
 (19) and (20), we obtain
 \begin{eqnarray}
 M_{A}(r) =&&4\pi
 \sigma_{0}\int_0^r[\frac{f^2}{r^2}+\frac{f'^2}{2A}+\frac{\lambda^2}{4}(f^2-1)^2+\rho_{q}]r^{2}dr\nonumber\\
 &&-4\pi
 \sigma_{0}r-\frac{4\pi \sigma_{0}\rho_{0}}{3w_{q}}r^{-3w_{q}}.
 \end{eqnarray}
 which is the effective mass function of global monopole. A global
 monopole solution should have $\lim_{r \rightarrow \infty} f = 1$.
 If this convergence is quick enough in flat space, $M_{A}(r)$ will also converge fast to a finite value. The effective
 mass of the solution is determined by the asymptotic value $\lim_{r \rightarrow r_h} M_{A}(r) \equiv
 M_{A}$. Numerical calculations show that the shape of $f(r)$ is quite
 insensitive to $\epsilon^{2}$ in the range $0 \leq \epsilon \leq
 1$, therefore the above discussion is consistent. In order to solve
 the system of equations uniquely, we need to introduce boundary
 conditions as follows
 \begin{eqnarray}
 f(0) = 0, A^{-1}(r_h) = B(r_h) = 0, A(0) = B(0) = 1.
 \end{eqnarray}\\
\\
\noindent {\it 4.2. A special example : $w_{q} = -\frac{2}{3}$
monopole}\\
\\
\noindent \indent As a characteristic example, we are going to
investigate $w_{q} =
 -\frac{2}{3}$ global monopole. Let us first discuss the asymptotic
 behavior at the origin. Expanding the functions we have
\begin{widetext}
 \begin{eqnarray}
 &&f(r\ll1)=f_{1}r+\frac{3}{8}\epsilon^{2}f_{1}\rho_{0}r^{2}+O(r^{3}),\rho_{q}(r\ll1)=\rho_{0}r^{-1}+O(r^{3}),\nonumber\\
 &&A^{-1}(r\ll1)=1-\frac{\epsilon^{2}\rho_{0}}{2}r-\frac{1}{12}(\epsilon^{2}\lambda^{2}+6\epsilon^{2}f_1^{2})r^{2}+O(r^{3}),\nonumber\\
 &&B(r\ll1)=1-\frac{\epsilon^{2}\rho_{0}}{2}r-\frac{1}{12}\epsilon^{2}\lambda^{2}r^{2}+O(r^{3}).
 \end{eqnarray}
 \end{widetext}
where $f_{1}$ is a free parameter to be determined numerically. The
asymptotic behavior $(r\gg1)$ is given by
\begin{widetext}
\begin{eqnarray}
&&f(r\gg1)=1-\frac{1}{\lambda^{2}}r^{-2}+\frac{2\epsilon^{2}-3}{2\lambda^{4}}r^{-4}-[\frac{\epsilon^{2}\rho_{0}(2\epsilon^{2}-3)}{\lambda^{6}}+\frac{2M_{A}}{\lambda^{4}}]r^{-5}+O(r^{-6}),\nonumber\\
&&A^{-1}(r\gg1)=1-\epsilon^{2}-\frac{\epsilon^2\rho_{0}}{2}r+M_{A}
r^{-1}-\frac{\epsilon^{2}}{\lambda^{2}}r^{-2}-\frac{\epsilon^4\rho_{0}}{2\lambda^4}r^{-3}+O(r^{-4}),\nonumber\\
&&B(r\gg1)=1-\epsilon^{2}-\frac{\epsilon^{2}\rho_{0}}{2}r+M_{A}
r^{-1}-\frac{\epsilon^2}{\lambda^2}r^{-2}+O(r^{-4}),\nonumber\\
&&\rho(r\gg1)=\rho_{0}r^{-1}+O(r^{-3}).
\end{eqnarray}
\end{widetext}
where the effective mass $M_{A}=\lim_{r \rightarrow r_h}M_{A}(r)$ is determined numerically.\\
\indent To go beyond the simplified model, we find out the
gravitational field of global monopole by numerically solving the
combined Einstein's and Goldstone field equations. In all numerical
calculations, we can choose $\lambda = 1$ without loosing
universality. The resulting solutions for $f(r)$ and $\rho(r)$ are
shown in Fig. 1, where the parameters are chosen as $\epsilon =
10^{-2}$, $w_{q} = -\frac{2}{3}$ and $\rho_{0} = 0, 0.1, 1$. It is
easily found that the solution for the Goldstone scalar field $f(r)$
is not much different from the one with $\rho_{0} = 0$. In the
$\rho_{0} = 0$ case, Eqs. (4) and (8)-(10) are reduced to the case
previously studied in great detail in \cite{Barriola, Shi,
Liebling}. We have repeated the calculations and confirmed the
previous results. From Fig. 1, we find that the shape of the curve
$f(r)$ is quite insensitive to the value of $\rho_{0}$ in the
interval $0 \leq \rho_{0} \leq 1$ not only asymptotically, but also
close to the origin. Notice that the shape of $f(r)$ is also
insensitive to the value of $\epsilon$ in the interval $0 \leq
\epsilon \leq 1$. However, the shape of $f(r)$ is sensitive if
$w_{q} \neq -\frac{2}{3}$ and $\rho_{0} > 1$ (see Figs. 3-4).
\begin{figure}
\epsfig{file=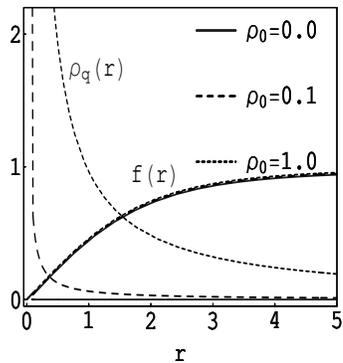,height=2.1in,width=2.5in}\caption{The
functions $f(r)$ and $\rho(r)$ are plotted vs the dimensionless
coordinate $r$ for different values of $\rho_{0}$ in the interval $0
\leq \rho_{0} \leq 1$. The parameters are chosen as $\epsilon =
10^{-2}$ and $w_{q} = -\frac{2}{3}$. The shape of the curve $f(r)$
is quite insensitive to the value of $\rho_{0}$. However, it is not
universal character for the cases with $w_{q} \neq -\frac{2}{3}$.}
\end{figure}
\begin{figure}
\epsfig{file=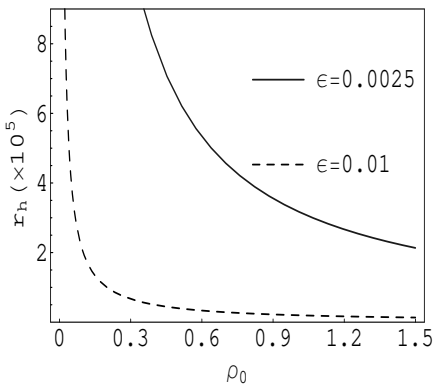,height=2.1in,width=2.5in}\caption{The value
of the radial dimensionless coordinate, where $A^{-1}(r=r_{h}) = 0$
is shown as function of $\rho_{0}$ for two different values of
$\epsilon$ and $\lambda = 1$.}
\end{figure}\\
\indent Using Eq. (14) we see that the global monopole surrounded
 by quintessence-like matter generates the outer horizon of de Sitter kind at $r =
 r_{h}$ if $-1 < w_{q} < -\frac{1}{3}$. Especially, the outer
 horizon $r_{h} =
 \frac{1-\epsilon^2}{\epsilon^2\rho_{0}}+\sqrt{\frac{(1-\epsilon^2)^2}{\epsilon^4\rho_0^2}+\frac{8}{3\lambda\rho_{0}}}$
 for $w_{q} = -\frac{2}{3}$ monopole in the simplified model. In the
 realistic model, we need to have the aid of numerical calculation.
 To investigate the appearance of horizon in the $w_{q} =
 -\frac{2}{3}$ monopole, we have fixed $\epsilon = 0.0025$ and
 $0.01$, respectively, and studied the value of
 zero of $A^{-1}(r)$, $r_{h}$ with $A^{-1}(r_{h}) = 0$ in dependence
 on $\rho_{0}$. Our results have been shown in Fig. 2. In the limit $\rho_{0} \rightarrow
 0$, the horizon tends to infinity. Obviously, $r_{h}$ is a
 decreasing function of $\rho_{0}$. For increasing $\rho_{0}$, the
 value of $r_{h}$ tends to zero as is demonstrated in Fig. 2.\\
 \\
 \noindent {\it 4.3. The limit of $w_{q} \rightarrow -\frac{1}{3}$}\\
 \\
\noindent \indent It is worthwhile to note that $w_{q} =
-\frac{1}{3}$ is a special
 case. Its asymptotic behavior must be discussed independently. In the
 simplified model, the interior and exterior solutions are reduced
 to, respectively
 \begin{eqnarray}
 ds^2=&&[1-\epsilon^2\rho_{0}-\frac{\epsilon^2\lambda^2}{12}r^2]dt^2-\frac{dr^2}{[1-\epsilon^2\rho_{0}-\frac{\epsilon^2\lambda^2}{12}r^2]}\nonumber\\&&-r^2(d\theta^2+\sin^2{\theta}
 d\varphi^2).
 \end{eqnarray}
and
\begin{eqnarray}
ds^2=&&[1-\epsilon^2(1+\rho_{0})-\frac{2G\sigma_{0}M}{r}]dt^2\\&&-\frac{dr^2}{[1-\epsilon^2(1+\rho_{0})-\frac{2G\sigma_{0}M}{r}]}-r^2(d\theta^2+\sin^2{\theta}
 d\varphi^2)\nonumber.
 \end{eqnarray}
 In the limit of $r \rightarrow \infty$, Eq. (27) implies that the
 metric component $|g_{rr}|$ tends to a constant value
 $[1-\epsilon^2(1+\rho_{0})]^{-1}$. Therefore, we can rescale the
 definition of radius by $r \rightarrow r/\sqrt{|g_{rr}|}$ so that
 the sphere surface gets the deficit solid angle since its area
 becomes $4\pi r^2/|g_{rr}|$ instead of $4 \pi r^2$. Certainly,
 Eq. (27) is not an exact solution for the metric around a $w_{q} =
 -\frac{1}{3}$ global monopole. Unfortunately, the exterior solution
 Eq. (27) is not consistent with the principal features of the realistic case $w_{q} = -\frac{1}{3}$,
 as we shall confirm in the text below.\\
 \indent Indeed, we can rigorously prove the non-existence of $w_{q} =
 -\frac{1}{3}$ global monopole by {\itshape reductio ad absurdum}. If there
 exists an exact monopole solution with $w_{q} = -\frac{1}{3}$, we
 have $f(0) = 0, f'(0)\neq 0, f''(0) =0$ and $\rho_{q} > 0$. In the
 neighbor of origin, the coupled Einstein and scalar field equations
 can be reduced to
 \begin{eqnarray}
 \frac{f'}{A}&=&\frac{f}{r},\\
 1-\frac{1}{A}&=&\epsilon^2(f^2+\rho_{q}r^2),\\
 \frac{A'}{A}&=&-\frac{B'}{B}.
\end{eqnarray}
From Eq. (28), we have $f = f_{1}r^{A(0)}$, where $f_{1}$ is an
integral constant. Since $f'(0) \neq 0$ and $f''(0) = 0$, we give
$A(0) = 1$ and $f_{1} = f'(0)$. Using this result, Eq. (29) can be
written as
\begin{equation}
f_{1}^2+\rho_{q} = 0.
\end{equation}
which conflicts with the condition $\rho_{q} > 0$.\\
\indent Note that we have employed the Lagrangian density (1) in
view of the above proof, where the Goldstone field and
quintessence-like matter are assumed to interact only through their
gravitational influence without direct interaction. Although
non-existence of $w_{q} = -\frac{1}{3}$ global monopole was
rigorously proved, the global monopole may exist if we introduce an
interactive term in the Lagrangian density (1).\\
\\
\noindent {\it 4.4. Asymptotic behavior}\\
\\
\indent In this subsection, we discuss the asymptotic behavior of
global monopoles for $-1 < w_{q} < -\frac{1}{3}$. At the origin, the
asymptotic behavior is
\begin{eqnarray}
f(r\ll1)&=&f_{1}r+O(r^{-3w_{q}}),\nonumber\\
A(r\ll1)&=&1-\frac{\epsilon^2\rho_{0}}{3w_{q}}r^{-3w_{q}-1}+O(r^{\beta}),\nonumber\\
B^{-1}(r\ll1)&=&1-\frac{\epsilon^2\rho_{0}}{3w_{q}}r^{-3w_{q}-1}+O(r^{\beta}),\nonumber\\
\rho_{q}(r\ll1)&=&\rho_{0}r^{-3w_{q}-3}+O(r^{\gamma}).
\end{eqnarray}
where
\begin{equation}
\beta =
\begin{cases}
2&\text{if $-1 < w_{q} < -\frac{2}{3}$},\\
-6w_{q}-2&\text{if $-\frac{2}{3} < w_{q} < -\frac{1}{3}$}.
\end{cases}
\end{equation}
and
\begin{equation}
\gamma=
\begin{cases}
-3w_{q}-1&\text{if $-1 < w_{q} < -\frac{2}{3}$},\\
-9w_{q}-5&\text{if $-\frac{2}{3} < w_{q} < -\frac{1}{3}$}.
\end{cases}
\end{equation}
The asymptotic behavior $(r\gg1)$ is given by
\begin{widetext}
\begin{eqnarray}
&&f(r\gg1)=1-\frac{1}{\lambda^2}r^{-2}-\frac{\epsilon^2\rho_{0}(2+3w_{q})}{3w_{q}\lambda^4}r^{-3w_{q}-5}+O(r^{-6w_{q}-8}),\nonumber\\
&&A^{-1}(r\gg1)=1-\epsilon^2
+\frac{\epsilon^2\rho_{0}}{3w_{q}}r^{-3w_{q}-1}+\frac{2G\sigma_{0}M_{A}}{r}+O(r^{-2}),\nonumber\\
&&B(r\gg1)=1-\epsilon^2
+\frac{\epsilon^2\rho_{0}}{3w_{q}}r^{-3w_{q}-1}+\frac{2G\sigma_{0}M_{A}}{r}+O(r^{-2}),\nonumber\\
&&\rho(r\gg1)=\rho_{0}r^{-3w_{q}-3}+O(r^{3w_{q}-1}).
\end{eqnarray}
\end{widetext}
\noindent {\it 4.5. Numerical results}\\
\\
\indent In this subsection, we still choose $\lambda^2 = 1$. As an
example of $w_{q} \neq -\frac{2}{3}$, the resulting solution for
$f(r)$ is shown in Fig. 3, where the parameters are chosen as
$\epsilon = 10^{-2}, w_{q} = -\frac{1}{2}$ and $\rho_{0} = 0,
10^{2}, 10^{3}$. It is easily found that the solution for the
Goldstone scalar field $f(r)$ is sensitive to the value of
$\rho_{0}$. It is inevitable that the disparity of $f(r)$ affects
the value of $M(r)$. The resulting solutions of $f(r)$ are shown in
Fig. 4 via different $w_{q}$, where other parameters are chosen as
$\epsilon = 0.01$ and $\rho_{0} = 10^{3}$. Notice that the shape of
$f(r)$ is sensitive to the value of $w_{q}$ except asymptotical
region. Obviously, the less the value of $w_{q}$ is, the more slowly
the curve of $f(r)$ tends to unit. The global monopole surrounded by
quintessence-like matter generates the horizon of de Sitter kind at
$r = r_{h}$ if $-1 < w_{q} < -\frac{1}{3}$. In Fig. 5, the value of
the radial dimensionless coordinate, where $A^{-1}(r=r_{h})=0$ is
shown as a function of $w_{q}$ for two different values of
$\epsilon$ and $\lambda = 1$. It is interesting to find that all the
outer horizons tend to infinite as $w_{q} \rightarrow -\frac{1}{3}$
in figure (5b). In Fig. 6, we plot $M(r)/\epsilon^{2}$ for different
values of $\rho_{0}$, where $M(r) \equiv M_{A}(r)/4\pi\sigma_{0}$,
the dimensionless parameter of the effective mass. Clearly, the mass
function is negative for all $r$. Especially, the shape of these
curves is considerably varied. In Fig. 7, we display
$M(r)/\epsilon^2$ for different values of $w_{q}$. Obviously, the
mass function is negative for all $r$, the value of which increases
with $w_{q}$.
\begin{figure}
\epsfig{file=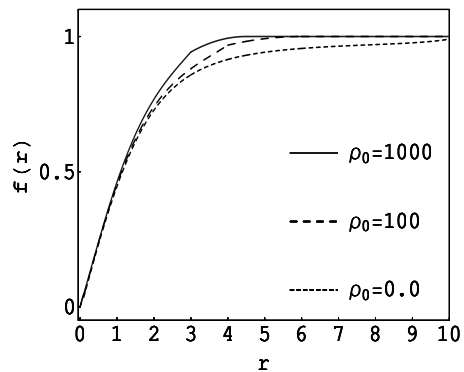,height=2.1in,width=2.5in}\caption{The
functions $f(r)$ are plotted vs the dimensionless coordinate $r$ for
different values of $\rho_{0}$. The parameters are chosen as
$\epsilon=10^{-2}, w_{q} = -\frac{1}{2}$ and $\lambda = 1$. The
shape of the curve $f(r)$ is sensitive to the value of $\rho_{0}$
except asymptotical region.}
\end{figure}
\begin{figure}
\epsfig{file=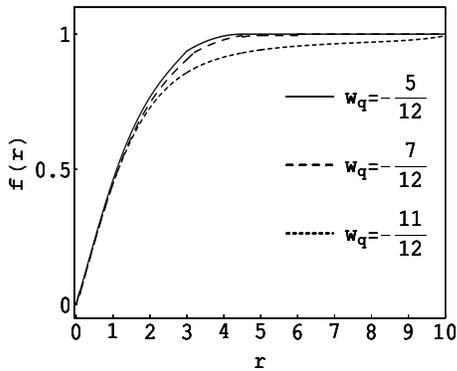,height=2.1in,width=2.5in}\caption{The
functions $f(r)$ are plotted vs the dimensionless coordinate $r$ for
different values of $w_{q}$. The parameters are chosen as
$\epsilon=10^{-2}, \rho_{0} = 10^{3}$ and $\lambda = 1$. The shape
of the curve $f(r)$ is sensitive to the value of $w_{q}$ except
asymptotical region.}
\end{figure}
\begin{figure}
\epsfig{file=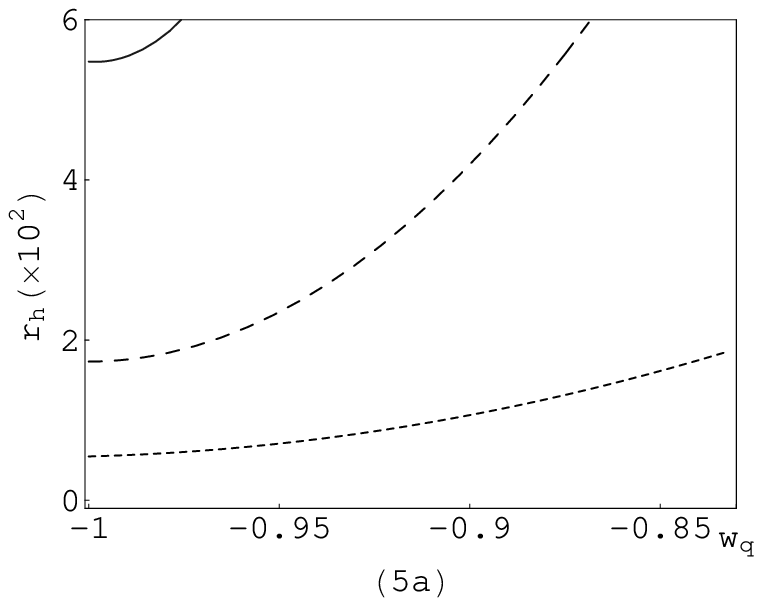,height=2.1in,width=2.5in}
\end{figure}
\begin{figure}
\epsfig{file=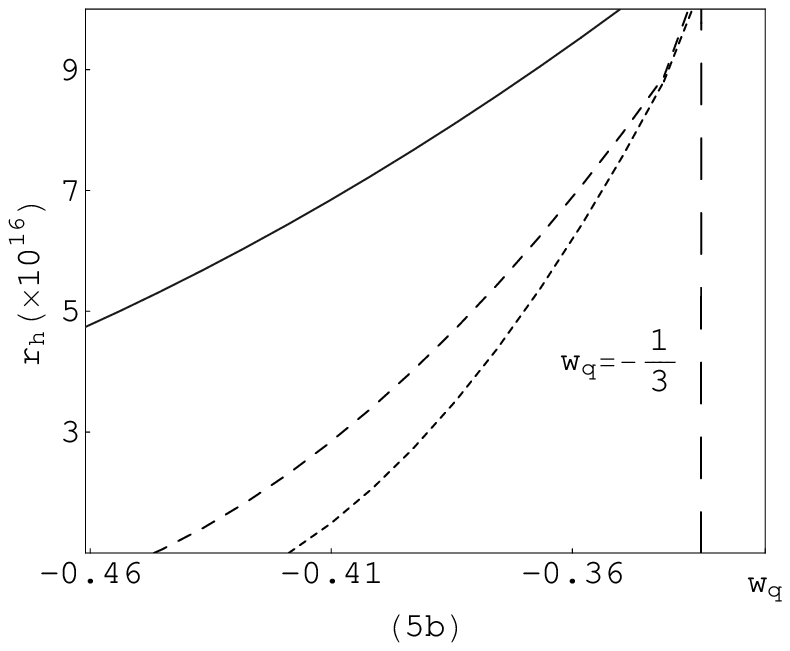,height=2.1in,width=2.5in}\caption{ The
horizon $r_{h}$ is shown as a function of $w_{q}$ for different
values of $\rho_{0} $, where solid line, long dash line and short
dash line correspond to $\rho_{0} = 0.1, 1, 10$ respectively. In
these figures, $r_{h}$ is plotted in $-1 \leq w_q \leq -0.83$ and
$-0.46 \leq w_q < -\frac{1}{3}$, respectively. All the outer
horizons tend to infinite as $w_{q} \rightarrow -\frac{1}{3}$.}
\end{figure}
\begin{figure}
\epsfig{file=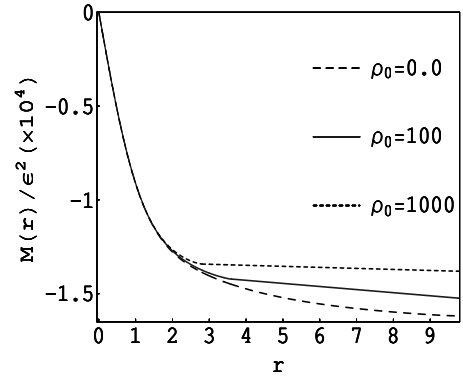,height=2.1in,width=2.5in}\caption{The mass
function $M(r)/\epsilon^2$ ($M(r) \equiv M_{A}(r)/4\pi\sigma_0$) is
plotted for different values of $\rho_{0}$. The other parameters are
$\epsilon = 10^{-2}$, $w_{q} = -\frac{1}{2}$ and $\lambda = 1$.}
\end{figure}
\begin{figure}
\epsfig{file=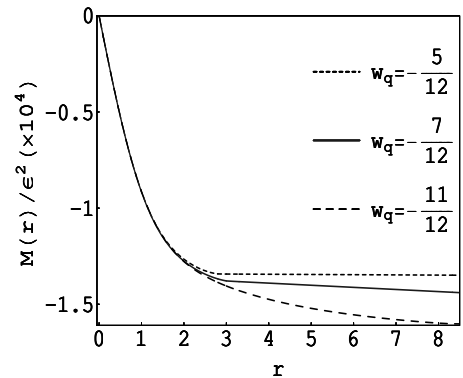,height=2.1in,width=2.5in}\caption{The mass
function $M(r)/\epsilon^2$ ($M(r) \equiv M_{A}(r)/4\pi\sigma_0$) is
plotted for different values of $w_{q}$. The other parameters are
$\epsilon = 10^{-2}$, $\rho_{0} = 10^{3}$ and $\lambda = 1$}
\end{figure}\\
\\
\noindent {\bf 5. Discussion and conclusions}\\
\\
\noindent \indent Topological defects are considered to be relevant
to structure formation in the early universe. Global defects are of
special interest in this context since they have a long-range scalar
field. This leads to the divergent mass in flat space, but renders a
strong gravitational effect when the global defects are investigated
in curved space. In the global monopole case, the linearly divergent
mass has an effective analogous to that of a deficit solid angle
plus a tiny mass at the origin. In this paper, we have got the
solutions of Einstein-scalar coupling equations for the static
spherically-symmetric quintessence-like matter surrounding a global
monopole, in which the scalar field and quintessence-like matter are
assumed to interact only through their gravitational influence
without
direct interaction. \\
\indent In a fundamental sense the stability of global monopole
arises from the nontrivial topology of the vacuum manifold
$\mathcal{M}$. The removal of the defect entails the infinite cost
of varying all of the field hence the defect is classically stable.
However, the global monopole was created in the early universe
therefore it might possibly  form thin shell like wormhole during
phase transitions. Rahaman and colleagues \cite{Rahaman} have
researched the thin shell wormhole in the context of global monopole
for Barriola and Vilenkin solution \cite{Barriola}. They have
obtained the time evolution equation of the radius $a$ of wormhole
throat. When the initial velocity $\dot{a}=0$, the radius is a
position of static equilibrium. They have also analyzed the
dynamical stability of the thin shell, considering linearized radial
perturbations around static solution of wormhole \cite{Rahaman}. By
applying this method, we can obtain a restriction on the model
parameters. In our models, the evolution equation of throat can be
written as
\begin{eqnarray}
&&\dot{a}+V(a)=0\\\
&&V(a)=V_{\textbf{BV}}(a)-\frac{\epsilon^2\rho_0}{2}a^{-3w_q-1}
\end{eqnarray}
\noindent where $V_{\textbf{BV}}(a)$ is the potential for
Barriola-Vilenkin case \cite{Rahaman}. By the linearization, the
stability of equilibrium configurations corresponds to the condition
$V''(a_0)>0$, where the prime denotes derivative with respect to
$a$. We have
\begin{equation}
V''(a)=V''_{\textbf{BV}}(a)-\frac 12(3w_q+1)(3w_q+2)\epsilon^2\rho_0
a^{-3w_q-3}
\end{equation}
\noindent for global monopole surrounded by quintessence-like
matter. In the $-\frac 23<w_q<-\frac 13$ case, we have
$V''(a)>V''_{\textbf{BV}}(a)$ so that the restriction of parameters
could be loosened; in the $-1<w_q<-\frac 23$ case, we have
$V''(a)<V''_{\textbf{BV}}(a)$ therefore the restriction becomes
tighter; in the $w_q=-\frac 23$ case, the restriction is the same as
Barriola-Vilenkin case.

\indent Our results are mainly as follows:\\
\indent(i) The global monopole solutions are possible under the
appropriate choice of internal parameter in the energy-momentum
tensor of quintessence-like matter, depending on the parameter of
equation of
state $-1 < w_{q} <-\frac{1}{3}$. \\
\indent(ii) In the $w_{q} = -\frac{1}{3}$ case, there exists the
solution of the static spherically-symmetric quintessence-like
matter surrounding a black hole. But the static
spherically-symmetric quintessence-like matter surrounding a global
monopole cannot exist in the $w_{q} \rightarrow -\frac{1}{3}$ limit,
because
$\rho_{q} \rightarrow 0$ if $w_{q} \rightarrow -\frac{1}{3}$.\\
\indent(iii) The distinguishing feature is the appearance of outer
horizon for the case of quintessence-like matter surrounding a
global monopole. The numerical results show that the radius of
horizon becomes larger and larger as the density of
quintessence-like matter decreases.\\
\indent(iv) In $w_{q} = -\frac{2}{3}$ case, Eq. (25) indicates that
$\rho_{0}$ is a higher order effect in contrast with $w_{q} \neq
-\frac{2}{3}$ case. Therefore, the basic features of $w_{q} =
-\frac{2}{3}$ monopoles are similar to ordinary global monopole if
we take $0 < \rho_{0} \leq 1$. However, the features of new global
monopole are considerably varied in $w_{q} \neq -\frac{2}{3}$ case.
Since current observations constrain  $-1.14 < w_{q} < -0.93$
\cite{D}, new global monopoles have interesting astrophysical
applications.\\
\\
\noindent {\bf Acknowledgments}\\
\\
\indent This work is supported in part by the National Nature
Science Foundation of China under grant No. 10473007 and grant No
10671128.\\



\end{document}